\documentclass[runningheads]{llncs}
\usepackage[T1]{fontenc}
\usepackage{graphicx}
\usepackage{hyperref}
\usepackage[utf8]{inputenc}
\usepackage{amsmath} 
\usepackage{amsfonts}
\usepackage{mathtools}
\usepackage{multirow}
\usepackage{booktabs}
\usepackage{setspace}
\usepackage{booktabs, multirow, xcolor, colortbl, array}

\usepackage{booktabs, colortbl, multirow, xcolor}
\definecolor{rowhl}{HTML}{EEF6FF}

\begin{document}
\title{Blasto-Net: An Explainable Multi-Task Learning for Blastocyst Segmentation, Grading, and Implantation Prediction}
%
%
\titlerunning{Blasto-Net: An Explainable Multi-Task Learning}

\author{Zahra Asghari Varzaneh \inst{1} \and
Reza Khoshkangini \inst{1} \and Magnus	Johnsson \inst{2} \and Thomas Ebner \inst{3} \and Lars Johansson \inst{4} 
}
\authorrunning{Z. Asghari Varzaneh et al.}
%
\institute{Department of Computer Science and Media Technology, Malmö University, Malmö, Sweden\\
\email{zahra.asghari-varzaneh@mau.se, reza.khoshkangini@mau.se} \and Research Environment of Computer Science, Kristianstad University, Sweden \and Kepler University Hospital, Krankenhausstr, Linz, Austria \and NewLifeAid-Global AB, Sweden}


%
\maketitle              
%

\begin{abstract}
This study introduces Blasto-Net, a multi-task deep learning Model for comprehensive blastocyst analysis. The proposed model performs three tasks simultaneously in a single forward pass: segmentation of the ZP, TE, and ICM compartments, morphological grading, and implantation outcome prediction. Accurate blastocyst analysis in in vitro fertilization (IVF) is challenging. The compartments often have similar textures
but very different structures. To address these challenges, Blasto-Net employs an EfficientNet-B3 encoder with a UNet-style decoder enhanced by the Convolutional Block Attention Module (CBAM) and a novel Edge-Aware Attention Module (EAAM), to effectively capture both semantic and boundary information.
To handle distinct compartment topologies, the network employs specialized segmentation heads and a composite region and boundary-based loss. Additionally, Grad-CAM++ visualizations have been adapted to verify the anatomical consistency of the model’s predictions.
Evaluated on a public HMC blastocyst dataset, Blasto-Net achieves Dice scores of 94.93\%, 91.60\%, and 88.82\% for ICM, ZP, and TE, respectively, alongside an implantation F1-score of 80.00\%. These results demonstrate that Blasto-Net offers an accurate, interpretable, and efficient solution for automated blastocyst assessment, with strong potential to support clinical decision-making in IVF.

\end{abstract}
\section{Introduction}
\begin{figure}[t]
    \centering
    \includegraphics[width=0.6\textwidth]{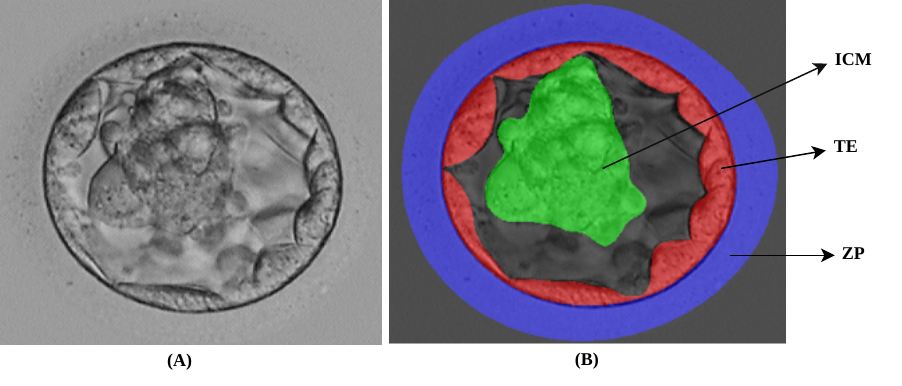}
    \caption{Microscopic image of a blastocyst (A) and a schematic representation (B) highlighting its three main components: ICM (green), TE (red), and ZP (blue).}
    \label{fig:blastocyst}
\end{figure}

Infertility represents a significant global issue that impacts around 17.5\% of adults worldwide, and its prevalence has grown over the past two decades~\cite{world2023infertility}. In vitro fertilization (IVF) is currently the most widely used assisted reproductive technology to address this problem. Despite significant advances, IVF success rates remain limited, and one of the key challenges is selecting the most viable embryo for transfer~\cite{gerris2005single,kromp2023annotated}. At the blastocyst stage, which is the final stage of development before transfer, embryo quality is assessed based on three distinct morphological compartments~\cite{balaban2000blastocyst}. The inner cell mass (ICM), which gives rise to the future fetus; the trophectoderm (TE), which mediates implantation and early endometrial interaction; and the zona pellucida (ZP),  the structural shell surrounding the embryo, whose thinning and stretching reflect the degree of blastocyst expansion. Throughout this study, we use the term ZP to refer to this expansion-related characteristic. Figure~\ref{fig:blastocyst} shows a blastocyst along with its three main components. The quality of these regions is strongly linked to implantation potential and pregnancy outcome~\cite{marikawa2009establishment}.
The Gardner grading system is the standard for blastocyst evaluation; however, grading through visual microscopic inspection is inherently subjective~\cite{kragh2019automatic}. Consequently, embryo scoring is prone to both inter- and intra-observer variability, which can compromise the consistency and reliability of morphological assessments. Therefore, these limitations highlight a strong need for objective, reproducible, and automated methods to support clinical decision-making~\cite{gardner2007analysis}.

Artificial intelligence, and especially deep learning, has shown strong potential in medical image analysis~\cite{zamani2025deep,varzaneh2025ensemble}. For embryo assessment, AI can provide key advantages over manual evaluation. It can extract detailed quantitative features from images, such as region boundaries, shape, texture, and spatial relations, which are difficult to measure manually but are potentially predictive of clinical outcomes~\cite{louis2021review}. 
For blastocyst analysis, segmentation, grading, and implantation prediction are inherently interconnected. Compartment morphology directly determines morphological grade, and the combined morphometric information from segmentation and grading provides the basis for implantation estimation. Given this interdependence, jointly optimizing these tasks within a multi-task learning model is a natural and well-motivated strategy that enables the model to learn richer shared representations than single-task approaches~\cite{caruana1997multitask}.
Another important challenge is the structural diversity of blastocyst components. While TE and ZP typically have ring-shaped structures, the ICM appears as a small, compact region. This makes ICM particularly difficult to segment due to its small size and class imbalance in images. Addressing these challenges requires models that can adapt to different structural characteristics simultaneously.\\
In this paper, we propose a multi-task learning (MTL) approach that simultaneously performs segmentation of ICM, TE, and ZP, predicts their morphological grades, and estimates implantation potential. The network uses an EfficientNet-B3 encoder paired with a UNet-style decoder. To improve feature selectivity at each level of the decoder, we incorporate Convolutional Block Attention Module (CBAM), which helps the network focus on the most relevant spatial regions and feature channels during decoding~\cite{woo2018cbam}. We also design topology-specific loss functions, use multi-scale edge maps as auxiliary supervision to sharpen boundary predictions, and integrate GradCAM++ visualization to make model decisions interpretable for clinical users.\\
The main contributions of this work are as follows:\\
- Proposing a single end-to-end multi-task network that simultaneously segments ZP, ICM, and TE, predicts morphological grades, and estimates implantation outcome.\\
- Integrating CBAM attention at every decoder level so that the network focuses on the most informative spatial regions and feature channels during feature reconstruction.\\
- Designing topology-aware loss functions for each compartment based on its geometry.\\
- Introducing multi-scale edge supervision derived from Sobel-Gaussian filtering to provide explicit boundary guidance and achieve sharper segmentation at inference. \\
- Employing GradCAM++ to provide visual interpretability and support trustworthy predictions.

\section{Related Work}
Accurate blastocyst assessment is a critical step in assisted reproductive technology, and various automated approaches have been introduced to improve embryo selection. Traditional evaluation methods are mainly based on morphological grading, which is subjective and may vary between experts~\cite{gardner1997culture}. To overcome these limitations, automated methods based on machine learning and deep learning have been proposed for more reliable analysis~\cite{hew2024artificial}. For instance, Saeedi et al.~\cite{saeedi2017automatic} proposed a method based on texture features, clustering, and watershed segmentation to detect TE and ICM, but it depended on manually designed features and was sensitive to image quality. In addition, Kheradmand et al.~\cite{kheradmand2016human} used a simple neural network based on DCT features for segmenting ZP, TE, and ICM. These methods depended on manual feature design and were not very robust.\\
More recent works use deep learning for more accurate segmentation. Rad et al.~\cite{rad2018multi} introduced BLAST-Net, which uses multi-scale context and progressive upsampling to better capture blastocyst structures. Arsalan et al.~\cite{arsalan2022detecting} proposed SSS-Net, using Sprint Convolutional Blocks with asymmetric kernel convolutions and depthwise separable convolutions.
The same group later proposed MASS-Net, combining four parallel downsampling branches at scales 1, 2, 4, and 8 with a Feature Booster Block that preserves spatial resolution to retain fine details of minor structures ~\cite{arsalan2022human}. Uysal et al.~\cite{uysal2022comparison} compared standard U-Net, Inception U-Net, ResUNet, and a custom Dilated Inception U-Net with group normalization and SELU activation for whole-embryo segmentation. Recently, Golfe et al.~\cite{golfe2025blastdiffusion} proposed BlastDiffusion to address the lack of annotated embryo data. Their method generates realistic synthetic oocyte images conditioned on developmental outcomes. Varzaneh et al.~\cite{varzaneh2025lightweight} developed a lightweight transformer-based method for predicting blastocyst formation. Their model combines DINOv2 for spatial feature extraction with an enhanced ViViT architecture for temporal sequence classification.\\
Despite these advances, most of these methods focus on segmentation only. MTL has been shown to improve generalization in medical imaging by forcing a shared encoder to serve multiple objectives simultaneously, and attention mechanisms such as CBAM have been used to enhance feature selection. In this study, we will apply both MTL and CBAM to address blastocyst segmentation, morphological grading, and implantation prediction from a single network.
\section{Methodology}
\vspace{-0.2cm}
\subsection {Overview of Blasto-Net}
\begin{figure}[t]
    \centering
    \includegraphics[width=1\textwidth]{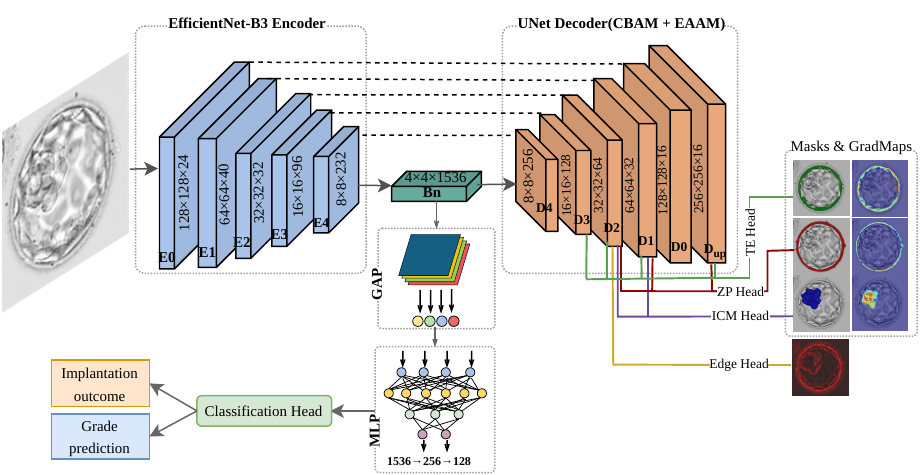}
    \caption{An overview of Blasto-Net architecture. The model follows an encoder-decoder structure where a shared EfficientNet-B3 encoder extracts hierarchical features, which are then processed by decoder blocks incorporating CBAM and EAAM attention. }
    \label{fig:overview}
\end{figure}

\vspace{-0.1cm}
We propose Blasto-Net, a multi-task approach that simultaneously solves three tasks from a single forward pass: pixel-level segmentation of three blastocyst compartments, prediction of morphological grades for each compartment, and binary implantation outcome prediction. As shown in Figure~\ref{fig:overview}, the process begins with a shared EfficientNet-B3 encoder, which extracts hierarchical feature representations from the input blastocyst image at multiple spatial resolutions (E0–E4). These features capture both low-level texture details and high-level semantic information necessary for downstream tasks. The encoded features are then reconstructed through a U-Net–style decoder. At each decoding stage (D4–D0), skip connections fuse encoder and decoder features to preserve spatial information. To enhance feature representation, each decoder block integrates Convolutional Block Attention Module (CBAM) for channel and spatial attention, along with the proposed Edge-Aware Attention Module (EAAM), which uses Sobel-based edge information to highlight boundaries and improve the delineation of compartment interfaces. From the final decoder features, the network branches into three topology-specific segmentation heads, each tailored to the anatomical characteristics of a blastocyst compartment. In parallel with the segmentation pathway, the deepest encoder features are used for classification, where global average pooling (GAP) is followed by a multi-layer perceptron (MLP) to extract high-level semantic information for predicting morphological grades and the binary implantation outcome. Meanwhile, multi-scale decoder features are selectively fused and routed to topology-specific segmentation heads for accurate segmentation of the ZP, ICM, and TE regions. Area ratios derived from the predicted masks are concatenated with the deep features to incorporate explicit morphological information. Through this coordinated pipeline, Blasto-Net enables effective information sharing among segmentation, grading, and implantation prediction tasks, leading to more accurate and clinically meaningful assessments. 

\subsection{Encoder: EfficientNet-B3 with Differential Feature Extraction}
The main body of the encoder is based on the EfficientNet-B3 architecture~\cite{tan2019efficientnet}. This architecture consists of six sequential stages, labeled as $\{s_0, s_1, s_2, s_3, s_4, bn\}$, and it generates feature maps with resolutions ranging from $H/2$ to $H/32$. The channel dimensions at each stage are as follows: 24, 40, 32, 96, 232, and 1536, respectively. EfficientNet-B3 is selected due to its effective scaling of depth, width, and resolution. It provides high-quality features while maintaining a moderate number of parameters.
\subsection {Decoder: UNet-Style with CBAM and EAAM}
\vspace{-0.1cm}
Each of the five DecoderBlock modules upsamples through a transposed convolution, concatenates the corresponding encoder skip connection, applies two $3\times3$ convolutions, and then sequentially applies CBAM and EAAM attention (Edge-Aware Attention Module)~\cite{wang2024edge}.
The CBAM reweights features through two cascaded gates. Channel attention uses both global average and global max pooling to produce per-channel scaling weights via a shared MLP. Spatial attention applies a $7 \times 7$ convolution over the concatenation of channel-wise average and max projections in Eq.~\eqref{eq:attn}:
\begin{equation}
\mathbf{F}' = \text{SpatialAttn}\!\left(\text{ChannelAttn}\!\left(\mathbf{F}\right)\right)
\label{eq:attn}
\end{equation}

EAAM extends standard attention by explicitly incorporating gradient information. Fixed depthwise Sobel kernels compute the gradient magnitude map $\mathbf{G}$ of the input feature, which is then used to gate the feature map through a learned $1\times1$ convolution as shown in Eq.~\eqref{eq:eaam}:

\begin{equation}
\text{EAAM}(\mathbf{F}) = \mathbf{F} + \mathbf{F} \odot \sigma\!\left(\text{Conv}_{1\times1}\!\left(\hat{\mathbf{G}}\right)\right)
\label{eq:eaam}
\end{equation}

where $\hat{\mathbf{G}} = \mathbf{G} / (\max \mathbf{G} + \varepsilon)$ is the normalised gradient magnitude and $\sigma$ is sigmoid. Because the Sobel kernels carry no learnable parameters, the raw edge signal flows directly from pixel gradients to the attention gate, providing boundary-sensitive feature modulation at every decoder level without adding trainable parameters beyond the gating convolution.

\subsection{Topology-Specific Segmentation Heads}
Instead of using one shared segmentation head for all compartments, we design three structure-aware branches that reflect the distinct morphological topology of each structure.
\vspace{-0.4cm} 
\subsubsection{ZP-Multi-Scale Ring Head:}
ZP presents as a thin outer ring. To capture both global ring structure and fine boundary details, the RingSegHead aggregates features from three decoder levels. As given in Eq.~\eqref{eq:zp} each feature is projected to 32 channels, upsampled to the output size, concatenated, and refined:
\begin{equation}
\hat{M}_{ZP} = \text{Conv}_{1\times1}\!\left(\text{ConvBlock}\!\left([\text{proj}(d_0) \| \uparrow\text{proj}(d_1) \| \uparrow\text{proj}(d_2)]\right)\right)
\label{eq:zp}
\end{equation}
\vspace{-0.7cm}
\subsubsection{ICM-LoG-Guided Refinement Branch:}
ICM is a small compact region, so it is more sensitive to class imbalance and confusion with TE. segmentation logit $\hat{M}_{ICM}$ is calculated in Eq.~\eqref{eq:ICM}.To preserve fine details, the branch uses decoder features from $d_2$ and $d_1$, and $\mathbf{h}$ is the fused feature from $d_2$ and $d_1$. Eq.~\eqref{eq:Log} is a Laplacian-of-Gaussian (LoG) map, extracted from the input image to guide the branch toward high-curvature regions:
\begin{equation}
\hat{M}_{ICM} = \text{out}\!\left(\mathbf{h} + \mathbf{h} \odot \sigma(\text{Conv}(\hat{E}_{LoG}))\right) 
\label{eq:ICM}
\end{equation}
\begin{equation}
E_{LoG} = |(\mathbf{x}_{gray} \star K_{LoG})|, \quad \hat{E}_{LoG} = E_{LoG} / (\max E_{LoG} + \varepsilon)
\label{eq:Log}
\end{equation}
The LoG guidance is injected through gated feature refinement via a sigmoid-activated $1\times1$ convolution, focusing the branch on high-curvature boundary regions where ICM is most likely to be found.
\vspace{-0.4cm}
\subsubsection{TE-Thin-Ring Head with Sobel Guidance:}
TE is the most challenging compartment because it is a thin irregular ring close to both ZP and ICM. The TE head aggregates four decoder levels $\{d_0, d_1, d_2, d_3\}$ and incorporates a Sobel edge map of the input image as explicit boundary guidance, shown in Eq.~\eqref{eq:sob}.
\begin{equation}
E_{Sob} = \sqrt{(\mathbf{x}_{gray} \star K_x)^2 + (\mathbf{x}_{gray} \star K_y)^2 + \varepsilon}
\label{eq:sob}
\end{equation}
where $(K_x)$ and $(K_y)$ are horizontal and vertical Sobel kernels, and (${x}_{\text{gray}}$) is the grayscale input.
The head outputs the segmentation logit $\hat{M}_{TE}$ and an auxiliary thin-ring (skeleton) logit $\hat{M}_{TE}^{thin}$. The latter is supervised using the morphological skeleton of the TE mask to enforce centerline representation and is discarded at inference.
\vspace{-0.3cm}
\subsection {Topology-Specific Multi-Task Loss Functions}
\vspace{-0.2cm}
The proposed segmentation loss $\mathcal{L}_{seg}$ is defined as a weighted composition of three topology-specific loss functions, $\mathcal{L}_{ZP}$, $\mathcal{L}_{ICM}$, and $\mathcal{L}_{TE}$, each designed for a specific anatomical compartment, as formulated in Equations~\eqref{eq:Lzp}, \eqref{eq:Licm}, and~\eqref{eq:Lte}:
\begin{align}
&\mathcal{L}_{ZP} = \alpha_{Lov}\mathcal{L}_{Lov} + (1{-}\alpha_{Lov})\mathcal{L}_{Dice} + 0.3\,\mathcal{L}_{DT} + \alpha_b\,\mathcal{L}_{bnd} + 0.8\,\mathcal{L}_{ring} \label{eq:Lzp} \\
&\mathcal{L}_{ICM} = 0.4\,\mathcal{L}_{Lov} + 0.3\,\mathcal{L}_{Tversky} + 0.2\,\mathcal{L}_{Focal} + 0.2\,\mathcal{L}_{bnd} \label{eq:Licm} \\
&\mathcal{L}_{TE} = 0.45\,\mathcal{L}_{Lov} + 0.15\,\mathcal{L}_{Dice} + 0.15\,\mathcal{L}_{bnd} + 1.2\,\mathcal{L}_{ring} + 0.4\,\mathcal{L}_{HD} + 0.25\,\mathcal{L}_{thin} \label{eq:Lte}
\end{align}
Here, $\mathcal{L}_{Lov}$ and Dice optimise overlap, while $\mathcal{L}{DT}$ and $\mathcal{L}{bnd}$ emphasise boundary accuracy via distance-based weighting. $\mathcal{L}{ring}$ enforces the hollow-ring topology of ZP/TE. Eq.~\eqref{eq:Licm}addresses severe class imbalance in the small-blob ICM region via $\mathcal{L}_{Tversky}$ with asymmetric $(\alpha=0.3, \beta=0.7)$, alongside $\mathcal{L}_{Focal}$ to focus on hard misclassified pixels. The loss $\mathcal{L}_{HD}$ captures boundary discrepancies, and $\mathcal{L}_{thin}$ supervises the TE skeleton.\\
Consequently, our proposed composite segmentation loss is defined as Eq.~\eqref{eq:seg}:

\begin{equation}
\mathcal{L}_{seg} = \frac{1.2\,\mathcal{L}_{ZP} + 1.5\,\mathcal{L}_{ICM} + 1.8\,\mathcal{L}_{TE}}{4.5} + 0.3\,\mathcal{L}_{ds} \label{eq:seg}
\end{equation}

where higher weights are assigned to the more challenging TE compartment, and $\mathcal{L}_{ds}$ denotes a deep supervision term.
\vspace{-0.1cm}
\subsection{Explainability via GradCAM++}
\vspace{-0.2cm}
To support clinical acceptance, we include GradCAM++ visualizations for each output~\cite{chattopadhay2018grad}. Hooks are registered on the first convolution of each topology-specific head and on the final encoder layer. The resulting activation maps highlight the exact image regions that most influenced each prediction, allowing clinical users to verify that the network is attending to anatomically correct compartments. This explainability capability is presented as a separate output alongside the segmentation masks.
\vspace{-0.2cm}
\section{Experimental Setup }
\vspace{-0.2cm}
\subsection{Dataset}
\vspace{-0.1cm}
In this study, we employed a publicly available blastocyst dataset consisting of 249 Hoffman Modulation Contrast (HMC) images. These images were collected from patients treated at the Pacific Centre for Reproductive Medicine (PCRM) in Canada between 2012 and 2016~\cite{saeedi2017automatic}. Each image is accompanied by a corresponding pixel-level ground truth annotation, annotated by expert embryologists. In addition to the segmentation masks, the dataset provides detailed grading information for major blastocyst components based on established clinical assessment criteria. This comprehensive annotation scheme enables a detailed assessment of both morphological structures and their quality.
\vspace{-0.2cm}
\subsection{Implementation details}
The dataset was split into training, validation, and test sets in a 70/15/15 ratio. Images were resized to 256×256 and normalized, while optional augmentations included flips, rotations, affine transforms, and Gaussian blur. The model was trained for 100 epochs with a batch size of 8, using the Adam optimizer with learning rate = 3e-4.
\vspace{-0.2cm}
\subsection{Evaluation Metrics}
Since the problem is multi-task, both segmentation and classification metrics were used. For segmentation tasks, Dice coefficient, Jaccard index (IoU), and boundary-aware losses were monitored~\cite{muller2022towards}. For classification tasks, accuracy, precision, recall, and F1-score were computed for each class~\cite{hossin2015review}. 

\section{Results and Discussion}

The proposed Blasto-Net model simultaneously addresses three clinically coupled tasks. 
\vspace{-0.2cm}
\subsection{Multi-Task Performance}
\begin{figure}[t]
    \centering
    \includegraphics[width=0.9\textwidth, 
                     trim=12.8cm 12cm 0cm 0.5cm, 
                     clip]{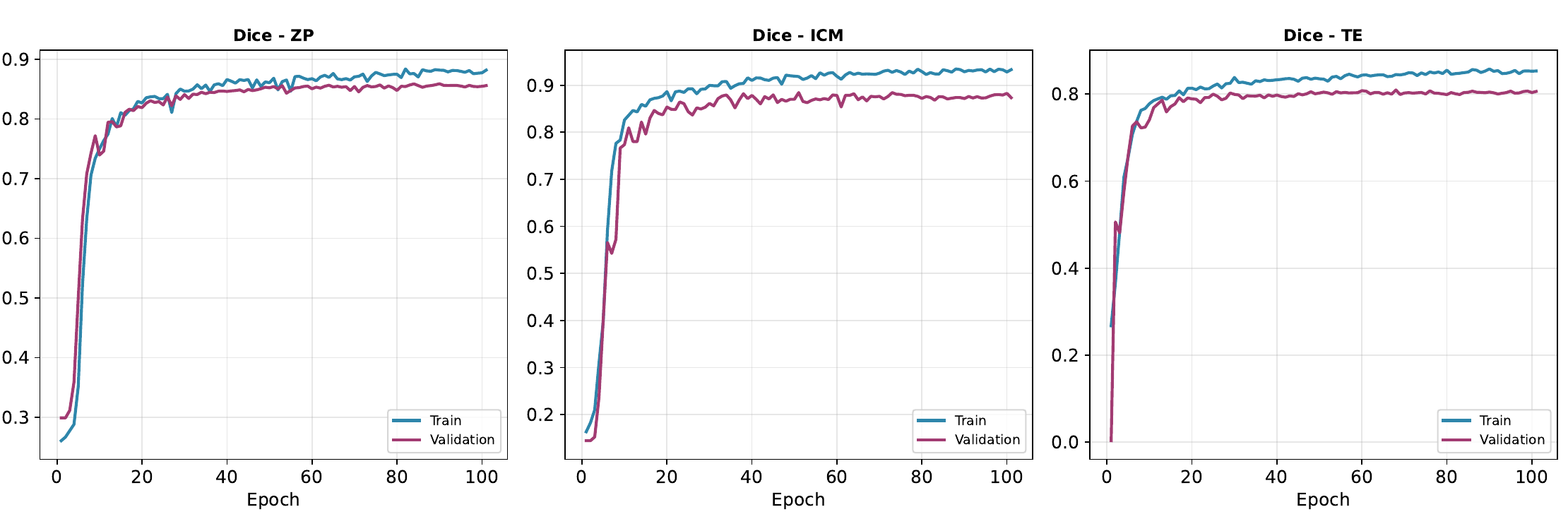}
    \caption{Dice curves over training epochs for each compartment (ZP, ICM, and TE) on training and validation data.}
    \label{fig:dice_curves}
\end{figure}

\begin{figure}[t]
    \centering
    \includegraphics[width=0.9\textwidth, 
                     trim=0cm 0cm 12.7cm 13.2cm, 
                     clip]{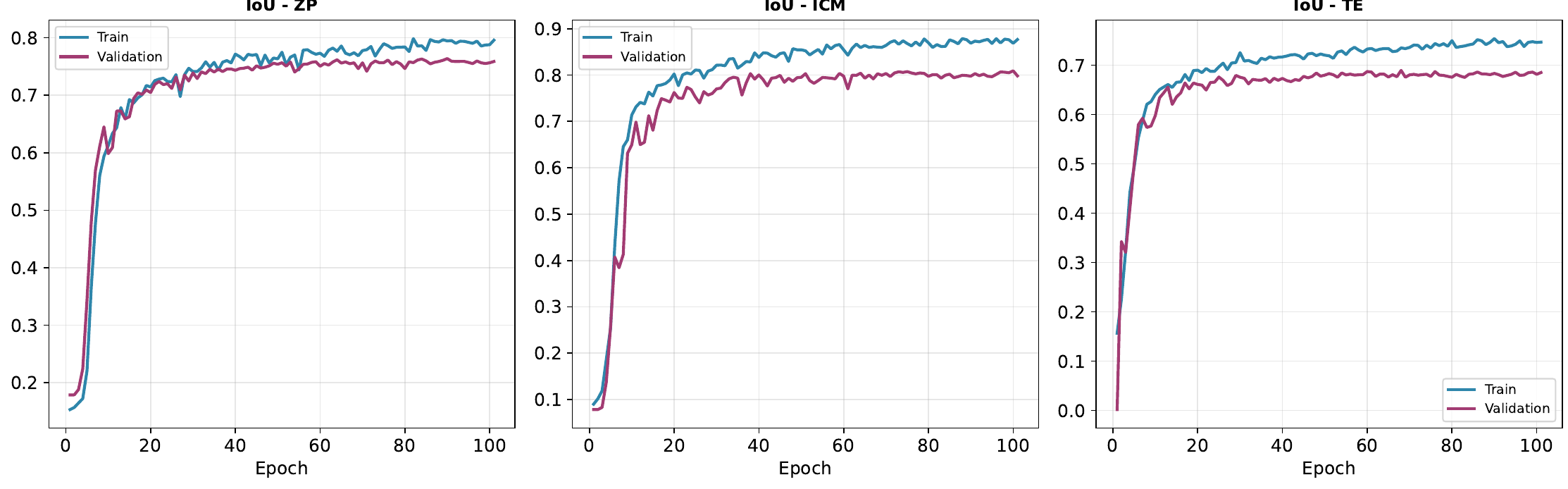}
    \caption{IoU curves over training epochs for each compartment (ZP, ICM, and TE) on training and validation data.}
    \label{fig:iou_curves}
\end{figure}

\begin{figure}[t]
    \centering
     \includegraphics[width=0.8\textwidth, 
                     trim=1cm 0cm 1cm 11cm, 
                     clip]{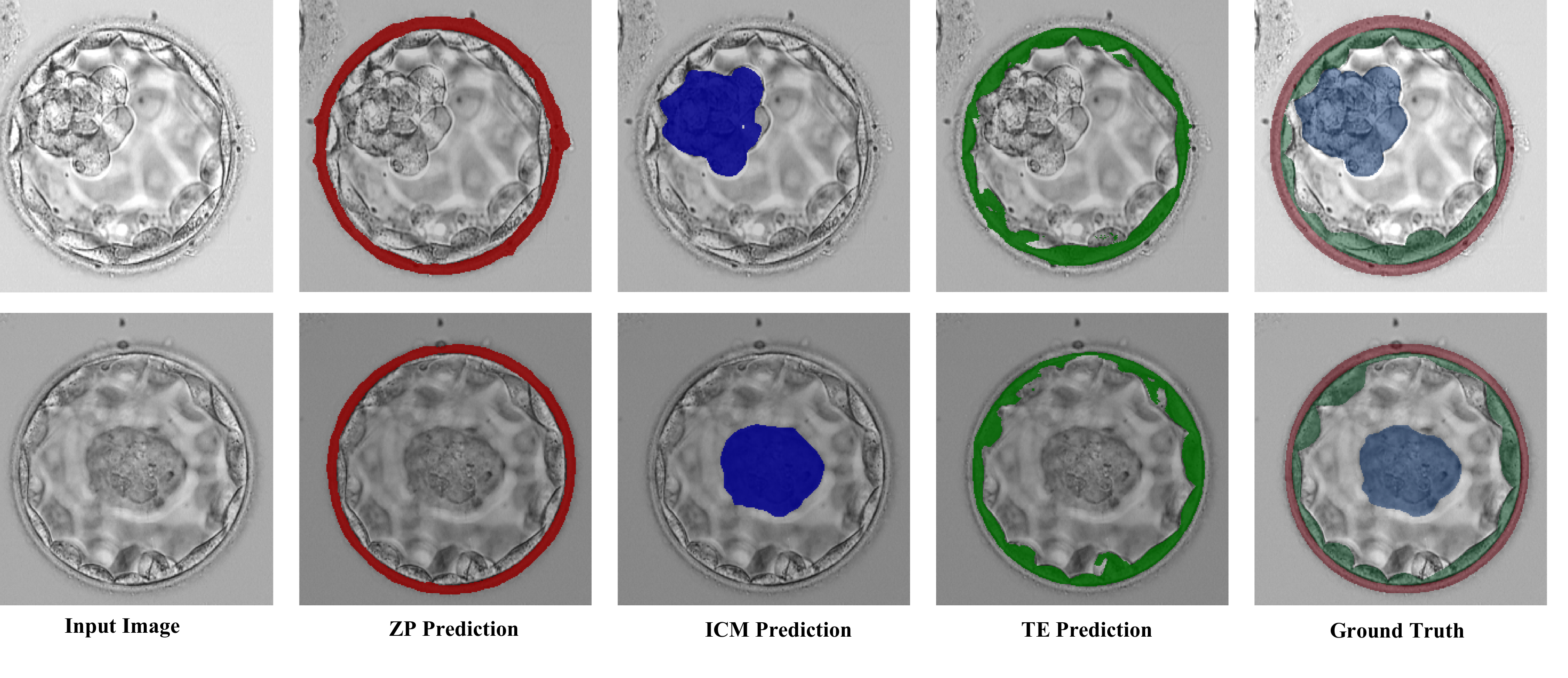}
    \caption{Predicted segmentation masks for all compartments on test samples.}
    \label{fig:seg_results}
\end{figure}

Since Blasto-Net is a multitask model, we evaluate it on three related clinical tasks: compartment segmentation, morphological grading, and implantation prediction. Table~\ref{tab:multitask_results} summarises all results together, which reflects the unified design of the proposed model.
The proposed model achieves strong segmentation performance across all three blastocyst compartments. 
ICM achieves the highest Dice score of 94.93\% and an IoU of 85.73\%. This result is notable because ICM is the smallest region and the most imbalanced region. ZP achieves a Dice of 91.60\% and 82.31\% IoU. TE, which presents the most difficult segmentation target due to its thin and irregular annular geometry, reaches 88.82\% Dice and 78.89\% IoU. The training progression of both Dice and IoU metrics across all three compartments is shown in Fig.~\ref{fig:dice_curves} and Fig.~\ref{fig:iou_curves}, respectively, where a consistent convergence pattern is observed for train and validation curves, with no sign of severe overfitting.
\begin{table*}[b]
\centering
\caption{Multi-task performance of Blasto-Net for segmentation and prediction tasks.}
\label{tab:multitask_results}
\renewcommand{\arraystretch}{1.15}
\setlength{\tabcolsep}{5pt}

\footnotesize

\begin{tabular}{lcc@{\hspace{12pt}}cccc}
\hline
\multirow{2}{*}{\textbf{Region / Task}} 
& \multicolumn{2}{c@{\hspace{24pt}}}{\textbf{Segmentation}} 
& \multicolumn{4}{c}{\textbf{Prediction/Grading Performance}} \\
\cline{2-7}
& Dice& IoU 
& Precision & Recall & F1-score & Accuracy \\
\hline

TE  
& 88.82 & 78.89 
& 89.47 & 69.12 & 75.00 & 76.00 \\

ICM 
& 94.93 & 85.73 
& 57.99 & 67.42 & 56.14 & 68.00 \\

ZP
& 91.60 & 82.31 
& 49.05 & 55.97 & 49.52 & 48.00 \\

\hline

\textbf{Implantation}  
& -- & -- 
& 69.57 & 94.12 & 80.00 & 76.00 \\

\hline
\end{tabular}
\end{table*}
Example segmentation results for two test samples are shown in Fig.~\ref{fig:seg_results}. Each row includes the input image, the predicted masks for the three compartments (ZP, ICM, and TE), and the ground-truth masks in the last column. The results show that the model correctly identifies all three structures and keeps their expected shapes, including the thin ring of ZP, the compact ICM region, and the irregular thin ring of TE.

For morphological grading, the dataset is imbalanced across all three grading tasks, which is common in clinical data. Despite this, the model shows reasonable performance. TE grading achieves the highest accuracy of 76\% with an F1-score of 75.00\%, while ICM grading reaches 68\% accuracy. Expansion (ZP) grading is more difficult task, with an accuracy of 48\% and F1 of 49.52\%, which is attributable to the higher ambiguity of this label in the dataset and the relatively small number of training samples per class. To reduce the effect of imbalance, we use label smoothing and class weighting. Overall, the results remain competitive given the limited data.
For implantation prediction, the model achieves strong results. It reaches a recall of 94.12\%, meaning most embryos that implanted are correctly identified. The F1-score of 80.00\% shows a good balance between precision and recall. This performance comes from the morphology-aware implantation head, which uses both learned features and area ratios between the predicted compartments. This helps the model make more biologically meaningful predictions

\vspace{-0.2cm}
\subsection{Edge Detection and Boundary Supervision}
\begin{figure}[t]
    \centering
    \includegraphics[width=0.4\textwidth, 
                     trim=0cm 14cm 0cm 0.5cm, 
                     clip]{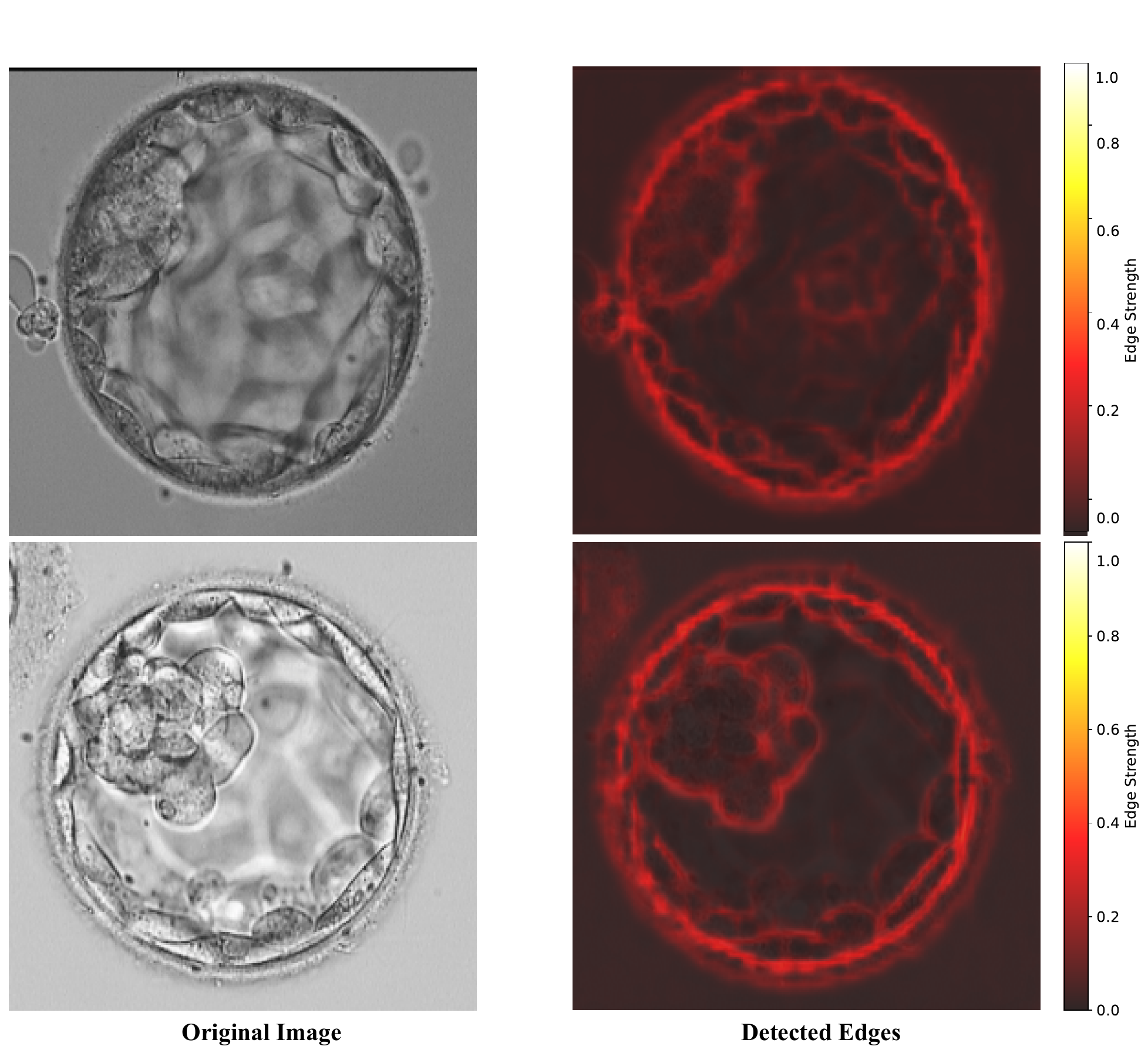}
    \caption{Auxiliary edge maps. Multi-scale boundary maps (Sobel–Gaussian) overlaid on the input image.}
    \label{fig:edge_detection}
\end{figure}

One important part of Blasto-Net is how it learns boundaries. During training, we use edge maps generated from Sobel and Laplacian-of-Gaussian (LoG) filters, together with the Edge-Aware Attention Module (EAAM) in each decoder block. This helps the model focus on edges more clearly. We use this approach because most segmentation errors happen at the boundaries, especially for the thin rings of ZP and TE. As shown in Fig.~\ref{fig:edge_detection}, the predicted edge maps for two test samples demonstrate the model’s capability to accurately delineate both the outer boundary of the ZP and the inner ring of the TE, even when these structures are closely adjacent.
\vspace{-0.3cm}
\subsection{Explainability via GradCAM++}
\begin{figure}[t]
    \centering
    \includegraphics[width=0.9\textwidth, 
                     trim=0cm 25cm 0cm 0cm, 
                     clip]{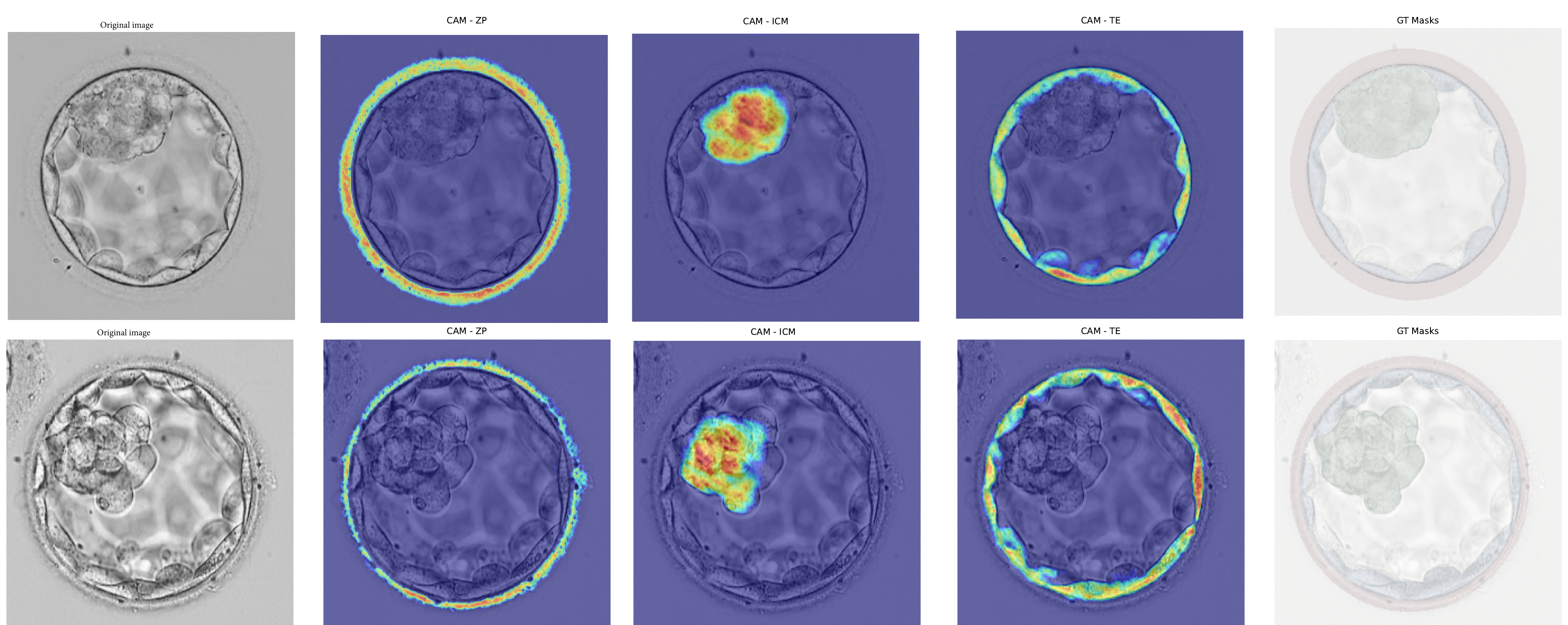}
    \caption{GradCAM++ explainability maps for segmentation }
    \label{fig:gradcam}
\end{figure}

\vspace{-0.1cm}
To support clinical interpretability and to validate that the network attends to anatomically correct regions, we integrate GradCAM++ visualizations for each segmentation output. We chose GradCAM++ because it produces sharper and more precise maps compared to standard GradCAM, which makes it more suitable for thin structures like ZP and TE. Fig.~\ref{fig:gradcam} shows these activation maps for two test samples. As you can see, the model focuses on the correct areas. For ZP, the attention is mainly on the outer ring. For ICM, it focuses on the compact cell region. For TE, it highlights the inner layer between the ZP and the blastocoel. Overall, these results show that the model is not just picking up background patterns, but is actually learning meaningful, task-related features. They also show that each segmentation head has learned to focus on its own target structure, even though they all share the same encoder–decoder backbone.

\vspace{-0.2cm}
\subsection{Comparison with the State-of-the-Art methods}
\vspace{-0.2cm}
Table~\ref{tab:seg_comparison} compares Blasto-Net with nine previously published methods on the same blastocyst dataset using the IoU metric. It is important to note that all prior methods address segmentation as a standalone task, without any concurrent grading or implantation prediction. Blasto-Net is the first model to address all three clinical objectives simultaneously within a single multi-task architecture. In terms of segmentation performance, Blasto-Net achieves competitive or superior results across all three compartments. For ICM, it achieves the highest IoU of 85.73\%, surpassing the previous best result from SSS-Net (Residual) at 84.94\%. This is particularly notable given that Blasto-Net is not focused on segmentation alone, but is also learning grading and implantation prediction at the same time. For TE, Blasto-Net reaches an IoU of 78.89\%, which is very close to TransFuse (78.10\%) and higher than all convolution-based baselines. For ZP, the model achieves 82.31\% IoU. This is comparable to transformer-based methods, although slightly below SSS-Net (Dense), which reports 84.51\%. This small gap on ZP is expected. SSS-Net is a single-task model, so it can dedicate its full capacity to segmentation. In contrast, Blasto-Net has to balance multiple objectives at once. Even so, it maintains strong and consistent performance across all compartments.
\begin{table*}
\centering
\caption{Comparison of Blasto-Net with the state-of-the-art segmentation methods using IoU (\%) metric.}
\label{tab:seg_comparison}
\small
\renewcommand{\arraystretch}{1.1}
\setlength{\tabcolsep}{6pt}
\begin{tabular}{lccc}
\hline
\textbf{Method} & \textbf{ZP IoU(\%)} & \textbf{TE IoU (\%)} & \textbf{ICM IoU (\%)} \\
\hline
U-Net (baseline) \cite{ronneberger2015u} & 79.32 & 75.06 & 79.03 \\
Ternaus U-Net \cite{iglovikov2018ternausnet} & 80.24 & 76.16 & 77.58 \\
DeepLab-V3 \cite{zhao2017pyramid} & 80.84 & 73.98 & 80.60 \\
UNet++ \cite{zhou2018unet++} & 78.43 & 74.23 & 79.65 \\
BlastNet \cite{rad2019blast} & 81.15 & 76.52 & 81.07 \\
SSS-Net (Residual) \cite{arsalan2022detecting} & \textbf{82.88} & 77.40 & 84.94 \\
SSS-Net (Dense) \cite{arsalan2022detecting} & \textbf{84.51} & 78.15 & 84.50 \\
TransFuse \cite{zhang2021transfuse} & 82.23 & 78.10 & 84.25 \\
TransUNet \cite{chen2021transunet} & 81.62 & 75.95 & 83.85 \\
\hline
\textbf{Blasto-Net (Ours)} & \textbf{82.31} & \textbf{78.89} & \textbf{85.73} \\
\hline
\end{tabular}
\end{table*}
\subsection{Ablation Study}
Table~\ref{tab:ablation} shows a step-by-step ablation study, where each component is added on top of the previous model. Edge supervision (M\textsubscript{1}) improves all three segmentation classes by 0.4–1.3\% Dice. This shows that providing boundary information helps the model better distinguish different structures. When the grade classification loss is introduced (M\textsubscript{2}), segmentation improves noticeably (+3.6\% for ZP, +3.7\% for ICM, and +2.3\% for TE). This indicates that learning morphological grades alongside segmentation helps the model build more informative features. Using topology-specific heads (M\textsubscript{3}) brings the biggest jump in implantation accuracy (+8.0\%), showing that adapting the decoder to each structure is particularly helpful for clinical prediction, even though it slightly reduces performance on ZP and ICM. This drop is addressed in the next step (M\textsubscript{4}), where topology-specific losses recover and further improve segmentation (+0.6–0.7\%). Adding deep supervision (M\textsubscript{5}) gives another small boost across all segmentation classes (around 0.4–1.3\%). Finally, the full model (M\textsubscript{6}), with the morphology-aware implantation head, achieves the strongest overall results: +5.8\% for TE, +2.5\% for ICM, and +11.0\% in implantation accuracy. This final setup performs best across all metrics, showing that each component plays a useful and complementary role.

\begin{table}[t]
  \centering
  \caption{Ablation study of Blasto-Net. Starting from an EfficientNet-B3 + UNet decoder
           baseline trained with Dice loss only (M\textsubscript{0}), each row
           incrementally adds one component. Percentage deltas
           ($\textcolor[HTML]{1D7A40}{\uparrow}$ /
            $\textcolor[HTML]{A32D2D}{\downarrow}$)
           show the change relative to the immediately preceding variant.
           \textbf{Bold} denotes the best result per column.}
  \label{tab:ablation}
  \setlength{\tabcolsep}{4pt}
  \footnotesize
  \begin{tabular}{@{}l ccc c@{}}
    \toprule
    \textbf{Model variant} &
    \multicolumn{3}{c}{\textbf{Dice (\%)}} &
    \textbf{Impl Acc (\%)} \\
    \cmidrule(lr){2-4}
    & ZP & ICM & TE & \\
    \midrule
    M\textsubscript{0}\;Baseline
      & 83.6 & 86.2 & 81.5 & 68.0 \\
    \midrule
    M\textsubscript{1}\;+Edge supervision
      & 84.0 \textcolor[HTML]{1D7A40}{\scriptsize$\uparrow$0.4}
      & 87.5 \textcolor[HTML]{1D7A40}{\scriptsize$\uparrow$1.3}
      & 82.3 \textcolor[HTML]{1D7A40}{\scriptsize$\uparrow$0.8}
      & 64.0 \textcolor[HTML]{A32D2D}{\scriptsize$\downarrow$4.0} \\
    M\textsubscript{2}\;+Grade loss
      & 87.6 \textcolor[HTML]{1D7A40}{\scriptsize$\uparrow$3.6}
      & 91.2 \textcolor[HTML]{1D7A40}{\scriptsize$\uparrow$3.7}
      & 84.6 \textcolor[HTML]{1D7A40}{\scriptsize$\uparrow$2.3}
      & 68.0 \textcolor[HTML]{1D7A40}{\scriptsize$\uparrow$4.0} \\
    M\textsubscript{3}\;+Topology-specific heads
      & 86.8 \textcolor[HTML]{A32D2D}{\scriptsize$\downarrow$0.8}
      & 90.7 \textcolor[HTML]{A32D2D}{\scriptsize$\downarrow$0.5}
      & 84.8 \textcolor[HTML]{1D7A40}{\scriptsize$\uparrow$0.2}
      & \textbf{76.0} \textcolor[HTML]{1D7A40}{\scriptsize$\uparrow$8.0} \\
    M\textsubscript{4}\;+Topology-specific losses
      & 87.4 \textcolor[HTML]{1D7A40}{\scriptsize$\uparrow$0.6}
      & 91.4 \textcolor[HTML]{1D7A40}{\scriptsize$\uparrow$0.7}
      & 85.4 \textcolor[HTML]{1D7A40}{\scriptsize$\uparrow$0.6}
      & 72.0 \textcolor[HTML]{A32D2D}{\scriptsize$\downarrow$4.0} \\
    M\textsubscript{5}\;+Deep supervision
      & 88.7 \textcolor[HTML]{1D7A40}{\scriptsize$\uparrow$1.3}
      & 92.4 \textcolor[HTML]{1D7A40}{\scriptsize$\uparrow$1.0}
      & 85.8 \textcolor[HTML]{1D7A40}{\scriptsize$\uparrow$0.4}
      & 65.0 \textcolor[HTML]{A32D2D}{\scriptsize$\downarrow$7.0} \\
    \rowcolor[HTML]{EEF6FF}
    M\textsubscript{6}\;Full model (ours)
      & \textbf{88.8} \textcolor[HTML]{1D7A40}{\scriptsize$\uparrow$0.1}
      & \textbf{94.9} \textcolor[HTML]{1D7A40}{\scriptsize$\uparrow$2.5}
      & \textbf{91.6} \textcolor[HTML]{1D7A40}{\scriptsize$\uparrow$5.8}
      & \textbf{76.0} \textcolor[HTML]{1D7A40}{\scriptsize$\uparrow$11.0} \\
    \bottomrule
  \end{tabular}
\end{table}

\section{Conclusion and future work}
\vspace{-0.2cm}
In this work, we introduced Blasto-Net, a unified multi-task framework for blastocyst analysis in IVF. The model simultaneously performs compartment segmentation, morphological grading, and implantation prediction from a single shared encoder–decoder backbone. By learning these tasks together, it models richer biological relationships than single-task approaches.
The architecture combines topology-specific segmentation heads, edge-aware attention, and a morphology-aware implantation module. A topology-specific loss strategy improves performance. Overall, these components address key challenges such as structural variation between compartments, class imbalance in the small ICM region, and thin-ring boundary precision for ZP and TE. GradCAM++ is also integrated to provide interpretable and anatomically meaningful predictions.
Experiments show strong performance across all tasks. Blasto-Net achieves state-of-the-art ICM IoU of 85.73\%, while also producing grading and implantation predictions, while these tasks not previously addressed together. The high implantation recall (94.12\%) is particularly important in IVF, where missing a viable embryo is costly.
Future work may extend the model to time-lapse embryo videos and incorporate patient-specific clinical data for more personalized predictions.

\begin{credits}
\subsubsection{\ackname} This work was supported by Vinnova, the Swedish Governmental Agency for Innovation Systems [Grant No. 2024-01462]. The funding source had no role in the design, execution, or publication of the study.

\end{credits}
%
%
%
%

\end{document}